\documentstyle[12pt]{article}
\newcommand{\bi}{\bibitem}
\begin{document}

\title {\Large\bf Duality and Bosonisation in Arbitrary Dimensions}
\vspace{1in}
\author {
{\bf R. Banerjee}\\
{\normalsize S.N.Bose National Centre for Basic Sciences.}\\
{\normalsize DB 17, Sector 1, Salt Lake City,
Calcutta 700064, INDIA}\\ }
\date{}
\maketitle

\noindent{\bf Abstract}

A functional integral approach is developed to discuss the
bosonisation of the massive Thirring and the massive Schwinger
models in arbitrary D-dimensions. It is found that these models,
to {\it all} orders in the inverse fermi mass, bosonise to a
theory involving a usual gauge field and a (D-2) rank
antisymmetric (Kalb-Ramond) tensor field. Explicit bosonisation
identities for the fermion current are deduced. Specialising to
the lowest order reveals (for any $D \geq 4$)
a mapping between the massive Thirring
model and the Proca model. It also establishes an exact duality
between the Proca model and the massive (D-2) rank Kalb-Ramond
model. Schwinger terms in the current algebra are computed.
Conventional bosonisation results in D=2, 3 are  reproduced.

\newpage
\section{Introduction}
The technique of bosonisation which consists in expressing a theory
of fermions in terms of bosons provides a powerful nonperturbative
tool for investigations in either quantum field theory [1] or condensed
matter systems [2]. This idea of bosonisation has been recently
extended in an interesting series of papers [3, 4] which reveal that the
fermionic and bosonic versions of the theory are two extremes of a
characterisation which, at intermediate stages, is a mixed
representation of apparently interacting fermions and bosons. It is
important to stress, however, that bosonisation (or its recent
extension) is well established in only two (i.e. 1+1) space-time
dimensions. This is because Schwinger terms which can give a
clue to bosonisation are rather complicated in higher
dimensions. Moreover, extracting bosonisation from a seemingly
interacting theory of bosons and fermions requires the
computation of the 1-cocycle for chiral transformations [5]. Apart
from the fact that such transformations are meaningful only in
even dimensions, it should also be realised that closed form
expressions for the 1-cocycle are readily calculable only in two
(1+1) space-time dimensions. Inspite of these difficulties, some
understanding of bosonisation in higher dimensions has been
attained [6-10]. Specially, in 2+1 dimensions, the bosonisation of the
massive Thirring model to leading order in the inverse fermion
mass has been performed [9]. The successful completion of this
program by extending the computations to all orders has also
been done [10]. In dimensions greater than 2+1, however, the
situation is rather obscure involving both technical and
conceptual problems. In [7], the bosonisation of a
massive  fermion interacting with an external potential has been
done but only upto the leading order in the inverse mass. The
bosonised theory, moreover, is found to be nonlocal. Ref[8], on
the other hand, considers the bosonisation of a free theory but
the analysis is once again valid only upto the leading order in
the inverse fermion mass.

In the present paper a functional integral approach is used to
systematically discuss the bosonisation of fermionic theories,
particularly the massive Thirring and Schwinger models, in any
dimensions. The bosonised versions as well as the various
bosonisation identities in the charge zero sector of the models
are derived to all orders in the inverse fermion mass. Contrary
to the findings reported in [7], all the boson-fermion
identifications are local. Specifically, it has been shown that
the massive Thirring model (MTM) in $D \ge 3$ space-time dimensions, to all
orders in the inverse mass, is equivalent to a gauge theory
involving a usual gauge field and a massless antisymmetric (D-2) rank
tensor field. Similar conclusions also hold for the
massive Schwinger model (MSM) except that the antisymmetric
tensor field is massive. Specialising next to the leading $(m^{-1})$
order it is shown that
for $D \ge 4$ the gauge theory representing the MTM
simplifies to the Proca theory.
Alternatively it can also be expressed in terms of a free massive
(D-2) rank antisymmetric tensor field (massive Kalb-Ramond
field). This approach to bosonisation therefore provides a
duality between the Proca field and the massive Kalb-Ramond
field in any dimensions $D \ge 4$. Incidentally the $D=3$
case may be recalled [9, 10] where the bosonisation of the MTM reproduces
the well known mapping [11] between the self-dual model of [12]
and the Maxwell-Chern-Simons theory [13]. This, as well as
the familiar $D=2$ dimensional bosonisation [1] are easily
reproduced. Finally, the leading order bosonisation of the
MSM just yields the conventional Maxwell theory.

In section 2 the MTM is expressed as a gauge theory of
apparently interacting fermions and bosons. This way of
interpreting the MTM is reminiscent of the approach in [3] where
the conventional two dimensional Thirring model is embedded in a
bigger gauge theory by performing a chiral transformation. The
role of gauge invariance and its connection to the dual (gauge
invariant) version of MTM is elaborated. A corresponding
construction for the MSM is also given. Section 3 discusses in
detail the bosonisation of MTM and MSM. Furthermore,
bosonisation identities mapping operators in the MTM and MSM
with corresponding operators in the dual (bosonised)theories are
given. The duality between the Proca model and the massive
Kalb-Ramond model for any $D \ge 4$ is revealed in section 4 as
a nontrivial application of the bosonisation program. As yet
another application the explicit computation of Schwinger terms,
to the leading $m^{-1}$ order, in the algebra of fermionic
currents is performed. Some
concluding remarks are presented in section 5.

\section{Dualisation of fermionic theories}
In this section it is first shown how the MTM can be expressed
in its dual form which turns out to be a gauge theory. Consider
the lagrangian in (D = d+1) space-time dimensions.
\footnote {The ensuing
analysis is for any $D \ge 3$. The D = 2 case will be treated
separately since it is special.}
\begin{equation}
{\cal L} = \bar{\psi} (i \partial\!\!\!/\, - m - \lambda
B\!\!\!\!/\, ) \psi + \frac{(-1)^{D}}{2(D-1)} F_{\mu_{1}
\mu_{2}....\mu_{D-1}} F^{\mu_{1} \mu_{2}....\mu_{D-1}} +
\epsilon_{\alpha\beta\mu_{1}\mu_{2}...\mu_{D-2}} B^{\alpha}
\partial^{\beta} A^{\mu_{1} \mu_{2}....\mu_{D-2}}
\end{equation}
where $A^{\mu_{1}\mu_{2}...\mu_{D}}$ is an antisymmetric D rank
tensor (the Kalb-Ramond field) and $F^{\mu_{1}....\mu_{D+1}}$ is
the corresponding field tensor,
\begin{equation}
F^{\mu_{1}...\mu_{D+1}} = \partial^{[\mu_{1}} A^{\mu_{2}....\mu_{D+1}]}
\end{equation}
with the symbol [] denoting antisymmetrisation. The other field
$B^{\mu}$ is an external vector field. This lagrangian is
invariant under the independent gauge transformations,
$$B_{\mu} \rightarrow B_{\mu} + \partial_{\mu} \omega ;\ \psi
\rightarrow e^{-i\lambda \omega} \psi$$
\begin{equation}
A_{\mu_{1}....\mu_{D}} \rightarrow A_{\mu_{1}...\mu_{D}} +
\partial_{[\mu_{1}} \Lambda_{\mu_{2}...\mu_{D}]}
\end{equation}
The equation of motion obtained by varying the distinct fields
in (1) are found to be,
\begin{equation}
(i\partial\!\!\!/\, - m - \lambda B\!\!\!\!/\,) \psi = 0
\end{equation}
\begin{equation}
\lambda j_{\mu} - \epsilon_{\mu\beta\mu_{1}...\mu_{D-2}}
\partial^{\beta} A^{\mu_{1}...\mu_{D-2}} = 0 ;\ j_{\mu} = \bar{\psi}
\gamma_{\mu}\psi
\end{equation}
\begin{equation}
\epsilon_{\alpha\beta\mu_{1}....\mu_{D-2}} \partial^{\beta}
B^{\alpha} + (-1)^{D} \partial^{\alpha}
F_{\alpha\mu_{1}...\mu_{D-2}} = 0
\end{equation}
Since $B_{\mu}$ is an external field it may be eliminated at the
classical level by using the equation (5). The dual
transformation now consists in making further use of (5) to
recast the field tensor in (1) in terms of the fermionic current,
\begin{equation}
\lambda^{2} j_{\mu} j^{\mu} =
\frac{(-1)^{D-1}(D-1)!}{(D-1)^{2}} F_{\mu_{1}...\mu_{D-1}}
F^{\mu_{1}...\mu_{D-1}}
\end{equation}
Using this result the lagrangian (1) simplifies to,
\begin{equation}
{\cal L}_{MTM} = \bar{\psi} (i \partial\!\!\!/\, - m) \psi -
\frac{\lambda^{2}}{2(D-2)!} j_{\mu} j^{\mu}
\end{equation}
which is just the lagrangian for the MTM. This is in fact the
dual version of (1).

To verify whether this duality (which was a classical result) is
preserved at the quantum level, it is necessary to work out the
partition function corresponding to (1),
$$Z = \int d[\psi, \bar{\psi}, B_{\mu},
A_{\mu_{1}...\mu_{D-2}}] \delta(\partial_{\mu}
B^{\mu})\delta(\partial_{\mu_{1}} A^{\mu_{1}...\mu_{D-2}})$$
\begin{equation}
{\em exp}\ i \int({\cal L} +
\epsilon_{\mu\nu\mu_{1}...\mu_{D-2}} \partial^{\nu}
A^{\mu_{1}...\mu_{D-2}} J^{\mu} + B_{\mu} K^{\mu})d^{D}x
\end{equation}
where external sources $J_{\mu}, K_{\mu}$ coupled to the
Kalb-Ramond field and $B_{\mu}$, respectively, have been
introduced. Moreover to preserve gauge invariance of the action,
the source $K_{\mu}$ must be conserved i.e. $\partial_{\mu}
K^{\mu} = 0$. The measure has been modified by inserting
$\delta$-functions as the (Lorentz) gauge fixing conditions
corresponding to the two independent gauge invariances (3). The
Gaussian integration over the Kalb-Ramond field is easily
performed by implementing the gauge $\partial_{\mu_{1}}
A^{\mu_{1}...\mu_{D-2}} = 0$ using 't Hooft's prescription to yield,
\begin{eqnarray}
Z &= &\int d[\psi,\bar{\psi}, B_{\mu}]
\delta(\partial_{\mu}B^{\mu}) {\em exp}\ i\int
d^{D}x[\bar{\psi}(i\partial\!\!\!/\, - m - \lambda
B\!\!\!\!/\,)\psi\nonumber \\
&  &+ \frac{(D-2)!}{2} (B_{\mu} + J_{\mu})^{2} + B_{\mu}
K^{\mu} - \frac{(D-2)!}{2} J_{\alpha}
\frac{\partial^{\alpha}\partial^{\beta}}{\Box} J_{\beta}]
\end{eqnarray}
Note the occurrence of a nonlocal term involving only the
sources. This will be cancelled after the $B_{\mu}$
intergration is done. To see this express
$\delta(\partial_{\mu} B^{\mu})$ as a Fourier transform with
variable $\beta(x)$. Then (10) may be written as,
\begin{eqnarray}
Z &= &\int d[\psi, \bar{\psi}, B_{\mu}, \beta] {\em exp}\
i\int d^{D}x[\bar{\psi}(i\partial\!\!\!/\, - m -
\lambda(B\!\!\!\!/\, + \frac{1}{(D-2)!} \partial\!\!\!/\,
\beta))\psi \nonumber \\
&  &+ \frac{(D-2)!}{2}\{( B_{\mu} + \frac{1}{(D-2)!}
\partial_{\mu}\beta)^{2} + J^{2}_{\mu} + 2J^{\mu}(B_{\mu} +
\frac{1}{(D-2)!} \partial_{\mu}\beta)\} \nonumber \\
&  &- \frac{1}{2(D-2)!}
\partial_{\mu}\beta\partial^{\mu}\beta -
J^{\mu}\partial_{\mu}\beta + (B_{\mu} + \frac{1}{(D-2)!}\partial_{\mu}\beta)
K^{\mu} \nonumber \\
&  &- \frac{(D-2)!}{2} J_{\alpha} \frac{\partial^{\alpha}
\partial^{\beta}}{\Box} J_{\beta}]
\end{eqnarray}
where use has been made of the fact that $K_{\mu}$ is a
conserved source and that the vector current $j_{\mu} =
\bar{\psi} \gamma_{\mu} \psi$ is conserved. In other workds I shall
always be considering some regularisation of the fermion
determinant which preserves gauge invariance thereby retaining
the classical conservation law $\partial_{\mu} j^{\mu} = 0$
obtainable from (5). Note that a guage invariant regularisation
also plays a key role in the recent discussions on bosonisation
given in [3-5]. Introducing the new fields $G_{\mu} = B_{\mu} +
\frac{1}{(D-2)!} \partial_{\mu}\beta$ and integrating over
$\beta$ yields,
\begin{eqnarray}
Z &= &\int d[\psi, \bar{\psi},  G_{\mu}] {\em exp}\ i\int
d^{D}x[\bar{\psi}(i\partial\!\!\!/\, - m - \lambda
G\!\!\!\!/\,)\psi \nonumber \\
&  &+ \frac{(D-2)!}{2} \{ G^{2}_{\mu} + J^{2}_{\mu} +
2J_{\mu}G^{\mu}\} + G_{\mu}K^{\mu}]
\end{eqnarray}
Note that, as announced earlier, the nonlocal term has been
precisely cancelled. Finally, integrating over $G_{\mu}$ leads to,
\begin{equation}
Z = \int d[\psi, \bar{\psi}]{\em exp}\ i\int
d^{D}x[\bar{\psi}(i\partial\!\!\!/\, - m)\psi -
\frac{\lambda^{2}}{2(D-2)!} j_{\mu}j^{\mu} +
\frac{\lambda}{(D-2)!} j_{\mu} ((D-2)! J^{\mu} + K^{\mu})]
\end{equation}
where $j_{\mu}$ is the fermionic current (5) and a
nonpropagating contact term has been dropped. Such terms will
henceforth always be ignored. In the absence of
sources it is seen that (13) represents the partition function
for the MTM. The normalisation of the current-current
interaction term also agrees with that obtained in (8) by a
classical analysis. I have thus shown that (1) represents the
lagrangian that is dual to the MTM. In the precise sense this
duality should be understood as an equivalence between the
partition functions (9) and (13). Furthermore, an inspection of
the source terms leads to the following mappings between the
Thirring current and the corresponding operators in the dual theory,
\begin{equation}
\lambda j_{\mu} \leftrightarrow
\epsilon_{\mu\nu\mu_{1}...\mu_{D-2}} \partial^{\nu}
A^{\mu_{1}...\mu_{D-2}} \leftrightarrow (D-2)! B_{\mu}
\end{equation}
These identifications are manifestations of the equations of
motion (5,6).

Likewise it is straightforward to construct the dual theory for
the MSM. Replacing the kinetic term in (1) by a mass term, we obtain,
\begin{equation}
{\cal L}' = \bar{\psi} (i\partial\!\!\!/\, - m - \lambda
B\!\!\!\!/\,) \psi - \frac{(-1)^{D}(D-2)!}{2}
A_{\mu_{1}...\mu_{D-2}} A^{\mu_{1}...\mu_{D-2}} +
\epsilon_{\alpha\beta\mu_{1}...\mu_{D-2}} B^{\alpha}
\partial^{\beta} A^{\mu_{1}...\mu_{D-2}}
\end{equation}
The classical equations of motion are once again given by (4),
(5) while (6) is modified to,
\begin{equation}
(-1)^{D}(D-2)! A_{\mu_{1}...\mu_{D-2}} +
\epsilon_{\alpha\beta\mu_{1}...\mu_{D-2}} \partial^{\beta}
B^{\alpha} = 0
\end{equation}
The dual version of (15) is now easily obtained by eliminating
the Kalb-Ramond field (which is no longer dynamical) using (16),
\begin{equation}
{\cal L}_{MSM} = \bar{\psi}(i \partial\!\!\!/\, - m - \lambda
B\!\!\!\!/\,)\psi - \frac{1}{4} B_{\mu\nu} B^{\mu\nu}
\end{equation}
where,
\begin{equation}
B_{\mu\nu} = \partial_{\mu} B_{\nu} - \partial_{\nu} B_{\mu}
\end{equation}
This is just the lagrangian for the MSM. A more formal derivation
follows by considering the partition function,
\begin{eqnarray}
Z' &= &\int d[\psi, \bar{\psi},  B_{\mu},
A_{\mu_{1}...\mu_{D-2}}] \delta(\partial_{\mu} B^{\mu}){\em exp}\
i\int({\cal L}' + \nonumber \\
&  &\epsilon_{\mu\nu\mu_{1}...\mu_{D-2}} \partial^{\nu}
A^{\mu_{1}...\mu_{D-2}} J^{\mu} + B_{\mu} K^{\mu})
\end{eqnarray}
where, as in the previous case,couplings with external sources
$J_{\mu}, K_{\mu}$ have been included. A Lorentz gauge fixing
delta function has been included in the measure. The second
delta function appearing in (9) is absent here because (15) is
no longer invariant under gauge transformations of the
Kalb-Ramond field. Performing the Gaussian integration over
$A_{\mu_{1}...\mu_{D-2}}$ yields,
\begin{eqnarray}
Z' &= &\int d[\psi, \bar{\psi},
B_{\mu}]\delta(\partial_{\mu} B^{\mu}){\em exp}\
i\int(\bar{\psi}(i\partial\!\!\!/\, - m - \lambda
B\!\!\!\!/\,)\psi \nonumber \\
&  &- \frac{1}{4} B^{2}_{\mu\nu} + \partial^{\nu} B_{\nu\mu}
J^{\mu} + B_{\mu} K^{\mu})
\end{eqnarray}
In the absence of sources, (20) represents the
partition function of the MSM in the Lorentz gauge. This
confirms the duality of (15) and (17). Comparison of the source
terms in (19) and (20) yields the identification between the
field strengths in the MSM and its dual version,
\begin{equation}
\partial^{\nu}B_{\nu\mu} \leftrightarrow
\epsilon_{\mu\nu\mu_{1}...\mu_{D-2}} \partial^{\nu} A^{\mu_{1}...\mu_{D-2}}
\end{equation}
which is also a manifestation of the equation of motion (16).

The analysis so far, as was particularly mentioned in the first
footnote  has been for any $D \geq 3$. For the special
case of D = 2, the lagrangian (1) gets replaced by,
\begin{equation}
{\cal L} = \bar{\psi}(i\partial\!\!\!/\, - m - \lambda
B\!\!\!\!/\,)\psi + \frac{1}{2}(\partial_{\mu}\theta)^{2} +
\epsilon_{\alpha\beta}B^{\alpha}\partial^{\beta}\theta
\end{equation}
The partition function in the presence of external sources is
given by,
\begin{equation}
Z = \int
d[\psi, \bar{\psi}, B_{\mu}, \theta]\delta(\partial_{\mu}B^{\mu}){\em
exp}\ i\int({\cal L} + \epsilon_{\mu\nu}\partial^{\nu}\theta
J^{\mu} + B_{\mu} K^{\mu})d^{2}x
\end{equation}
Contrary to the case $D\geq 3$ (see (9)), the partition function
for D = 2 has only one constraining delta function since the
lagrangian (22) is invariant under only one gauge transformation
(the first one in (3)). Following identical steps that led from
(9) to (13) enables one to carry out the integrations over the
Bose fields in (23) to yeild,
\begin{equation}
Z = \int d[\psi, \bar{\psi}]{\em exp}\ i\int
d^{2}x[\bar{\psi}(i\partial\!\!\!/\, - m)\psi -
\frac{\lambda^{2}}{2} j_{\mu} j^{\mu} + \lambda j_{\mu}(J^{\mu}
+ K^{\mu})]
\end{equation}
which is just the partition function for the MTM in the presence
of external sources. In fact it agrees with the general form
given in (13) for D = 2.

One can similarly show that in D = 2 the theory dual to the MSM
(17) is governed by the Lagrangian obtained by replacing the
Kalb-Ramond field in (15) by a scalar,
\begin{equation}
{\cal L}' = \bar{\psi}(i\partial\!\!\!/\, - m - \lambda
B\!\!\!\!/\,)\psi - \frac{1}{2} \theta^{2} +
\epsilon_{\alpha\beta} B^{\alpha} \partial^{\beta}\theta
\end{equation}

Having established the duality of both the MTM and MSM for any
$D \geq 2$ with their corresponding embedded versions, it is
straightforward to analyse the bosonisation of these models.
This will be presented in the next section.

\section{Bosonisation}
It is best to illustrate bosonisation by starting from the simplest
example which is in D = 2 and then proceeding to higher dimensions.

\noindent
{\bf i) D = 2 dimensions :}

Let me consider the massless version in which case the Thirring model
is known to be exactly solvable. The partition function follows from
(24),
\begin{equation}
Z = \int d[\psi, \bar{\psi}] exp\ i \int
d^{2}x[\bar{\psi}i\partial\!\!\!/\, \psi - \frac{\lambda^{2}}{2}
j_{\mu}j^{\mu} + \lambda j_{\mu}(J^{\mu} + K^{\mu})]
\end{equation}
and the dual (embedded) version is obtained from (22) and (23),
\begin{eqnarray}
Z &= &\int
d[\psi, \bar{\psi}, B_{\mu}, \theta]\delta(\partial_{\mu}B^{\mu})exp\
i \int[\bar{\psi}(i\partial\!\!\!/\, - \lambda B\!\!\!\!/\,)\psi +
\frac{1}{2}(\partial_{\mu}\theta)^{2} \nonumber \\
  &  &+ \epsilon_{\alpha\beta} B^{\alpha}\partial^{\beta}\theta +
  \epsilon_{\alpha\beta}\partial^{\beta}\theta J^{\alpha} +
  B_{\mu}K^{\mu}]
  \end{eqnarray}
To get the bosonised form of (26) the fermion integration in (27) is
first performed. This a well known [14] expression obtained from a gauge
invariant regularisation,
\begin{eqnarray}
Z &= &\int d[B_{\mu}, \theta]\delta(\partial_{\mu}B^{\mu})exp\ i
\int[-\frac{\lambda^{2}}{2\pi} B^{\mu}(g_{\mu\nu} -
\frac{\partial_{\mu}\partial_{\nu}}{\Box})B^{\nu} +
\frac{1}{2}(\partial_{\nu}\theta)^{2}\nonumber \\
  &  &+ \epsilon_{\alpha\beta}(B^{\alpha} +
  J^{\alpha})\partial^{\beta}\theta + B_{\mu}K^{\mu}]d^{2}x
  \end{eqnarray}
Implementing the gauge $\partial_{\mu}B^{\mu} = 0$ using 't Hooft's
prescription and carrying out the $B_{\mu}$-integration yields,
\begin{equation}
Z = \int d\theta exp\ i \int[\frac{1}{2}(1 +
\frac{\pi}{\lambda^{2}})(\partial_{\mu}\theta)^{2} +
\epsilon_{\mu\nu}\partial^{\nu}\theta(\frac{\pi}{\lambda^{2}} K^{\mu}
+ J^{\mu})]d^{2}x
\end{equation}
It is thus seen how the massless Thirring model gets identified with a
free massless scalar theory. Comparison of the source terms in (26)
and (29) reveal the mapping between the Thirring current and the
topological current in the scalar theory,
\begin{equation}
\lambda j_{\mu} \leftrightarrow N
\epsilon_{\mu\nu}\partial^{\nu}\theta
\end{equation}
modulo a normalisation N which is $\frac{\pi}{\lambda^{2}}$ or 1. This
is a manifestation of the familiar arbitrariness in the definition of
the Thirring current. For $\lambda^{2} = \pi$ both definitions agree
and the Thirring model maps identically to the massless free scalar
theory with,
\begin{equation}
j_{\mu} \leftrightarrow \frac{1}{\sqrt{\pi}}
\epsilon_{\mu\nu}\partial^{\nu}\theta;\
\bar{\psi}i\partial\!\!\!/\, \psi \leftrightarrow
\frac{1}{2}(\partial_{\mu}\theta)^{2}
\end{equation}
reproducing the well known identifications.

In a similar spirit consider the partition function for the massless
Schwinger model (20),
\begin{equation}
Z' = \int d[\psi, \bar{\psi}, B_{\mu}]\delta(\partial_{\mu}B^{\mu})exp\
i \int[\bar{\psi}(i\partial\!\!\!/\, - \lambda B\!\!\!\!/\,)\psi -
\frac{1}{4} B_{\mu\nu} B^{\mu\nu} + (\partial^{\nu}B_{\nu\mu})J^{\mu}]
\end{equation}
Its dual (embedded) version follows from (25) which is the
2-dimensional analogue of (19),
\begin{equation}
Z' = \int d[\psi, \bar{\psi},
B_{\mu},\theta]\delta(\partial_{\mu}B^{\mu})exp\ i
\int[\bar{\psi}(i\partial\!\!\!/\, - \lambda B\!\!\!\!/\,)\psi -
\frac{\theta^{2}}{2} + \epsilon_{\mu\nu} B^{\mu} \partial^{\nu}\theta
+ \epsilon_{\mu\nu} J^{\mu}\partial^{\nu}\theta]d^{2}x
\end{equation}
Doing the fermionic intergration yields,
\begin{equation}
Z' = \int d[B_{\mu}, \theta]\delta(\partial_{\mu}B^{\mu})exp\ i
\int[\frac{\lambda^{2}}{2\pi} B^{\mu}(g_{\mu\nu} -
\frac{\partial_{\mu}\partial_{\nu}}{\Box}) B^{\nu} -
\frac{\theta^{2}}{2} + \epsilon_{\mu\nu}(B^{\mu} +
J^{\mu})\partial^{\nu}\theta]d^{2}x
\end{equation}
Working out the $B_{\mu}$ integration by implementing the gauge
$\partial_{\mu}B^{\mu} = 0$ using 't Hooft's prescription leads to,
\begin{equation}
Z' = \int d\theta exp\ i
\int[\frac{\pi}{2\lambda^{2}}(\partial_{\mu}\theta)^{2} -
\frac{\theta^{2}}{2} +
\epsilon_{\mu\nu}J^{\mu}\partial^{\nu}\theta]d^{2}x
\end{equation}
Scaling $\theta \rightarrow \frac{\lambda}{\sqrt{\pi}}\theta$ gives
the desired structure,
\begin{equation}
Z' = \int d\theta exp\ i \int[\frac{1}{2}(\partial_{\mu}\theta)^{2} -
\frac{\lambda^{2}}{2\pi}\theta^{2} +
\frac{\lambda}{\sqrt{\pi}}\epsilon_{\mu\nu}J^{\mu}\partial^{\nu}\theta]d^{2}x
\end{equation}
This reproduces the result that the Schwinger model bosonises to a
massive free scalar theory with mass $M = \frac{\lambda}{\sqrt{\pi}}$.
The corresponding mapping between operators is given by,
\begin{equation}
\partial^{\nu}B_{\nu\mu} \leftrightarrow
M\epsilon_{\mu\nu}\partial^{\nu}\theta
\end{equation}

The analysis for the massive (either Thirring or Schwinger) models
follows by computing the fermion determinant as a perturbative
expansion about the massless case. Using the result of [15] one can
reobtain the known [1, 14] equivalence of the MTM(MSM) with the massless
(massive) sine- Gordon theory.

\noindent {\bf ii)D = 3 dimensions:}

As has been illustrated the bosonised version is obtained by
explicitly evaluating the fermion determinant. This is usually done by
using a gauge invariant regularisation and the expression is, in
general, non-local. Although an exact result does not exist, it is
possible to perform the computations under certain approximations. In
particular, an expansion in the large fermion mass limit is quite
popular. The 3-dimensional example is rather special since the leading
term in this expansion is given by the Chern-Simons three form. It
should also be mentioned that subsequent analysis is valid only in the
large mass limit. The partition function for the MTM follows from
(13),
\begin{equation}
Z = \int d[\psi,\bar{\psi}]exp\ i \int
d^{3}x[\bar{\psi}(i\partial\!\!\!/\, - m)\psi -
\frac{\lambda^{2}}{2}j_{\mu}j^{\mu} + \lambda j_{\mu}(J^{\mu} +
K^{\mu})]
\end{equation}
while its dual version is obtainable from (9),
\begin{eqnarray}
Z &= &\int
d[\psi,\bar{\psi},B_{\mu},A_{\mu}]\delta(\partial_{\mu}B^{\mu})\delta(\partial_{\mu}A^{\mu})exp\
i \int d^{3}x \nonumber \\
  &  &\bar{\psi}(i\partial\!\!\!/\, - m - \lambda B\!\!\!\!/\,)\psi -
\frac{1}{4}F_{\mu\nu} F^{\mu\nu} +
\epsilon_{\mu\nu\rho}B^{\mu}\partial^{\nu}A^{\rho} +
\epsilon_{\mu\nu\rho}
J^{\mu}\partial^{\nu}A^{\rho} + B_{\mu} K^{\mu}]
\end{eqnarray}
Computing the fermion determinant in (38) in inverse powers of m
[16],
\begin{eqnarray}
Z &= &\int
d[B_{\mu},A_{\mu}]\delta(\partial_{\mu}B^{\mu})\delta(\partial_{\mu}A^{\mu})exp\
i \int d^{3}x[- \frac{\lambda^{2}}{8\pi}\epsilon_{\mu\nu\lambda}
B^{\mu}\partial^{\nu}B^{\lambda}\nonumber \\
  &  &+\frac{\lambda^{2}}{24\pi m}B_{\mu\nu}B^{\mu\nu} + ..... -
  \frac{1}{4}F_{\mu\nu}F^{\mu\nu} +
  \epsilon_{\mu\nu\rho}B^{\mu}\partial^{\nu}A^{\rho} \nonumber \\
  &  &+ \epsilon_{\mu\nu\rho} J^{\mu}\partial^{\nu}A^{\rho} +
  B_{\mu}K^{\mu}]
\end{eqnarray}
where the dots are the higher order Seeley co-effecients in the
expansion of the fermion determinant. Thus the MTM, to all orders in
$m^{-1}$, can be mapped on to a gauge theory (40) with the
identifications,
\begin{equation}
\lambda j_{\mu} \leftrightarrow
\epsilon_{\mu\nu\rho}\partial^{\nu}A^{\rho} \leftrightarrow B_{\mu}
\end{equation}
The gauge theory (40) can be further simplified by performing the
Gaussian $A_{\mu}$ integration,
\begin{eqnarray}
Z &= &\int dB_{\mu}\delta(\partial_{\mu}B^{\mu})exp\ i \int d^{3}x[-
\frac{\lambda^{2}}{8\pi} \epsilon_{\mu\nu\lambda}
B^{\mu}\partial^{\nu}B^{\lambda} \nonumber \\
  &  &+\frac{\lambda^{2}}{24\pi m} B_{\mu\nu}B^{\mu\nu} + ..... +
  \frac{1}{2}(B_{\mu} + J_{\mu})^{2} + B_{\mu} K^{\mu} - \frac{1}{2}
  J_{\mu} \frac{\partial^{\mu}\partial^{\nu}}{\Box} J_{\nu}]
\end{eqnarray}
The nonlocal piece can be cancelled by following similar steps that
led from (10) to (12). Expressing $\delta(\partial_{\mu}B^{\mu})$ by
its Fourier transform with variable $\beta$ and then doing the $\beta$
integration yields,
\begin{equation}
Z = \int d G_{\mu} exp\ i \int d^{3}x[\frac{1}{2}G^{2}_{\mu} -
\frac{\lambda^{2}}{8\pi}\epsilon_{\mu\nu\lambda}
G^{\mu}\partial^{\nu}G^{\lambda} + \frac{\lambda^{2}}{24\pi m}
(\partial_{\mu}G_{\nu} - \partial_{\nu}G_{\mu})^{2} + .....+
G_{\mu}(J^{\mu} + K^{\mu})]
\end{equation}
where new fields $G_{\mu} = B_{\mu} + \partial_{\mu}\beta$ have been
introduced. Eq.(43) represents the complete bosonisation of the MTM.The
dots are to be identified with the dots occurring in (40) while the
Thirring current gets mapped with the basic field $G_{\mu}$,
\begin{equation}
\lambda j_{\mu} \leftrightarrow G_{\mu}
\end{equation}
Note that $G_{\mu}$, in contrast to $B_{\mu}$ is a gauge invariant
field. It is essentially the St\"uckelberg transformed $B_{\mu}$.

Coming next to the MSM, note that its partition function is given by
(20) while the dualised version is obtained from (19),
\begin{eqnarray}
Z' = \int d[\psi, \bar{\psi}, B_{\mu},
A_{\mu}]\delta(\partial_{\mu}B^{\mu})exp\ i \int
d^{3}x[\bar{\psi}(i\partial\!\!\!/\,& -& m -\lambda
B\!\!\!\!/\,)\psi  +
\frac{1}{2} A^{2}_{\mu} + \epsilon_{\alpha\beta\mu}
B^{\alpha}\partial^{\beta}A^{\mu}\nonumber\\ & +&
\epsilon_{\alpha\beta\mu}J^{\alpha}\partial^{\beta}A^{\mu} + B_{\mu}
K^{\mu}]
\end{eqnarray}
Computing the fermion determinant as done in going from (38) to (40),
\begin{eqnarray}
Z' &= &\int d[B_{\mu}, A_{\mu}]\delta(\partial_{\mu}B^{\mu})exp\ i \int
d^{3}x[- \frac
{\lambda^{2}}{8\pi}\epsilon_{\mu\nu\lambda}B^{\mu}\partial^{\nu}B^{\lambda}
+ \frac{\lambda^{2}}{24\pi m} B^{2}_{\mu\nu} + ......\nonumber \\
  &  &+ \frac{1}{2}A^{2}_{\mu}+\epsilon_{\alpha\beta\mu}(B^{\alpha} +
  J^{\alpha})\partial^{\beta}A^{\mu} + B_{\mu}K^{\mu}]
\end{eqnarray}
Doing the $A_{\mu}$ integration yeilds,
\begin{eqnarray}
Z' &= &\int dB_{\mu}\delta(\partial_{\mu}B^{\mu})exp\ i \int d^{3}x[-
\frac{\lambda^{2}}{8\pi}
\epsilon_{\mu\nu\lambda}B^{\mu}\partial^{\nu}B^{\lambda} \nonumber \\
  &  &+(\frac{\lambda^{2}}{24\pi m} - \frac{1}{4}) B^{2}_{\mu\nu} +
  .....+ J^{\mu}\partial^{\nu} B_{\nu\mu} + B^\mu K_\mu]
\end{eqnarray}
which gives the complete bosonised form of the MSM.

\noindent {\bf iii) $D \geq 4$ dimensions :}

The main difference from the 3-dimensional example is that the leading
term in the fermion determinant is a Maxwell piece instead of the
Chern-Simons 3-form. This leads to a bosonisation which is
(nontrivially) different from the 3-dimensional theory. Moreover,
contrary to a recent finding [7] which analyses the bosonisation of a
fermion coupled to an external potential in four dimensions, nonlocal
terms never appear in my discussion.

The partition function for the MTM $(D \geq 4)$ follows from (13),
while the dual version is given by (9). Computing the fermion
determinant in (9) using a gauge invariant regularisation yields,
\begin{eqnarray}
Z &= &\int d[B_{\mu},
A_{\mu_{1}....\mu_{D-2}}]\delta(\partial_{\mu}B^{\mu})\delta(\partial_{\mu_{1}}A^{\mu_{1}....\mu_{D-2}})exp\
i\int d^{D}x \nonumber \\
  &  &[- n_{D}\lambda^{2}B_{\mu}(\Box g^{\mu\nu} -
  \partial^{\mu}\partial^{\nu})B_{\nu} + ....+ \frac{(-1)^{D}}{2(D-1)}
  F_{\mu_{1}\mu_{2}...\mu_{D-1}} F^{\mu_{1}\mu_{2}...\mu_{D-1}}
  \nonumber \\
  &  &+ \epsilon_{\mu\nu\mu_{1}...\mu_{D-2}}
  B^{\mu}\partial^{\nu}A^{\mu_{1}...\mu_{D-2}}+
  \epsilon_{\mu\nu\mu_{1}...\mu_{D-2}}
  J^{\mu}\partial^{\nu}A^{\mu_{1}...\mu_{D-2}} + B_{\mu}K^{\mu}]
\end{eqnarray}
where only the leading term involving a dimension dependent
normalisation $n_{D}$ has been explicitly written. The subsequent dots
denote the other terms in the Seeley expansion of the fermion
determinant. This gives the complete bosonised form of the MTM, valid
to all orders in $m^{-1}$. The mapping between the Thirring current
and the relevant operators in the bosonised theory (48) is just given by
(14). It is now simple to perform the Gaussian integration over the
Kalb-Ramond field. This yields,
\begin{eqnarray}
Z = \int dB_{\mu}\delta(\partial_{\mu}B^{\mu})exp\ i \int
d^{D}x[n_{D}\lambda^{2}B^{\mu\nu}B_{\mu\nu}& +& .... +
\frac{(D-2)!}{2}(B_{\mu} + J_{\mu})^{2}\nonumber\\ & -& \frac{(D-2)!}{2}
J_{\mu}\frac{\partial^{\nu}\partial^{\nu}}{\Box}J_{\nu} +
B_{\mu}K^{\mu}]
\end{eqnarray}
As shown in earlier examples, the nonlocal term is eliminated by
expressing $\delta(\partial_{\mu}B^{\mu})$ by its Fourier transfer
with variable $\alpha(x)$ and then doing the integration over
$\alpha(x)$.It leads to a local result,
\begin{equation}
Z = \int d G_{\mu}exp\ i \int
d^{D}x[n_{D}\lambda^{2}(\partial_{\mu}G_{\nu} -
\partial_{\nu}G_{\mu})^{2}+....+ \frac{(D-2)!}{2} G^{2}_{\mu} + (D-2)!
G_{\mu}J^{\mu} + G_{\mu}K^{\mu}]
\end{equation}
where new fields $G_{\mu} = B_{\mu} + \frac{1}{(D-2)!}
\partial_{\mu}\alpha$, obtained from $B_{\mu}$ by a St\"uckelberg
transformation, were introduced. Eq.(50) gives another bosonised
representation, valid to all orders in the inverse mass, for the MTM.
The mapping between the Thirring current and the gauge invariant field
$G_{\mu}$ is obtained by comparing the source terms in (13) and (49),
respectively,
\begin{equation}
\lambda j_{\mu} \leftrightarrow (D-2)! G_{\mu}
\end{equation}
The final part of this section is devoted to discussing the MSM, whose
partition function is given in (20). Computing the fermion determinant
of its dual theory (19), one finds,
\begin{eqnarray}
Z' &= &\int d[B_{\mu},
A_{\mu_{1}....\mu_{D-2}}]\delta(\partial_{\mu}B^{\mu})exp\ i \int
d^{D}x [n_{D}\lambda^{2}B_{\mu\nu}B^{\mu\nu} +.....-
\frac{(-1)^{D}(D-2)!}{2} \nonumber \\
  &  &A^{2}_{\mu_{1}...\mu_{D-2}} +
  \epsilon_{\alpha\beta\mu_{1}...\mu_{D-2}} (B^{\alpha} +
  J^{\alpha})\partial^{\beta}A^{\mu_{1}...\mu_{D-2}} + B_{\mu} K^{\mu}]
\end{eqnarray}
which gives the bosonised form with the mapping (21). Performing the
integration over the Kalb-Ramond field yields,
\begin{equation}
Z' = \int dB_{\mu}\delta(\partial_{\mu}B^{\mu})exp\ i \int
d^{D}x[(n_{D}\lambda^{2} - \frac{1}{4})B_{\mu\nu}B^{\mu\nu} +
....+(\partial_{\mu}B^{\mu\nu})J_{\nu}+ B_{\mu}K^{\mu}]
\end{equation}
This result can also be obtained directly by computing the fermion
determinant from (20).

\section{Applications}
It is worthwhile to discuss some applications of the generalised
approach to bosonisation discussed in the preceding sections. The
analysis done uptill now (for $D\geq 3$) is valid to all orders in the
inverse fermi mass. Some interesting consequences follow if one
restricts to the leading (lowest) order term.

\noindent {\bf i) D = 3 space-time dimensions:}

The bosonised version of the MTM is given by (43). In the lowest (i.e.upto
$(m^{-1})$) order, this simplifies to,
\begin{equation}
Z = \int d G_{\mu} exp\ i \int d^{3}x[\frac{1}{2} G^{2}_{\mu} -
\frac{\lambda^{2}}{8\pi}\epsilon_{\mu\nu\lambda}G^{\mu}\partial^{\nu}G^{\lambda}+
G_{\mu}J^{\mu}]
\end{equation}
where one of the sources $K_{\mu}$ has been dropped since it provides
no additional information. In the absence of the source term, this is
the partition function of the self-dual model of [12]. This
correspondence between the MTM and the self dual model was obtained
earlier in [9, 10]. Note further that the mapping (44) remains unaltered.
The Thirring current, therefore, obeys a self-dual equation analogous
to $G_{\mu}$,
\begin{equation}
j_{\mu} - \frac{\lambda^{2}}{4\pi} \epsilon_{\mu\nu\lambda}
\partial^{\nu}j^{\lambda} = 0
\end{equation}

It is possible to give another bosonised form for the MTM by starting
from (40). Performing the $B_{\mu}$ integration leads to,
\begin{eqnarray}
Z &= &\int dA_{\mu}\delta(\partial_{\mu}A^{\mu})exp\ i \int d^{3}x
\nonumber \\
  &  &[\frac{2\pi}{\lambda^{2}}
  \epsilon_{\mu\nu\lambda}A^{\mu}\partial^{\nu}A^{\lambda} -
  \frac{1}{4} F_{\mu\nu}F^{\mu\nu} +
  \epsilon_{\mu\nu\rho}J^{\mu}\partial^{\nu}A^{\rho}]
\end{eqnarray}
In the absence of sources this is just the partition function for the
Maxwell-Chern-Simons (MCS) theory in the Lorentz gauge. The
equivalence of the MTM with this theory was previously explored
in [9, 10].
Comparing the source terms in (13) and (56) yields the mapping,
\begin{equation}
\lambda(j_{\mu})_{MTM} \leftrightarrow
(\epsilon_{\mu\nu\rho}\partial^{\nu}A^{\rho})_{MCS}
\end{equation}
Furthermore, since (54) and (56) are just dual descriptions of the
same theory, namely the MTM in the leading $m^{-1}$ expansion, these
theories must be completely equivalent. This, incidentally, reproduces
the well known [9-11] equivalence between the self-dual model (54) and the MCS
theory (56) with the identification,
\begin{equation}
(G_{\mu})_{Self-dual} \leftrightarrow (\epsilon_{\mu\nu\lambda}
\partial^{\nu}A^{\lambda})_{MCS}
\end{equation}
Coming next to the MSM, observe that its bosonised version is given by
(47). In the lowest order, this is the partition function for the MCS
theory in the Lorentz gauge.  But I have just shown that this theory is also
the bosonised
version of the MTM. Consequently, in the leading $m^{-1}$
approximation, the partiion functions for the MTM and the MSM become
equivalent.

\noindent {\bf ii) $D\geq 4$ space-time dimensions:}

The bosonised version of the MTM is given by (50). Considering the
leading term only and suitably scaling the $G_{\mu}$-field clearly
shows that the partition function corresponds to the massive Maxwell
model or the Proca model. Likewise, the Thirring
current gets mapped on to the Proca field by (51). Note furthermore
from (50) that the sources $J_{\mu}, K_{\mu}$ are coupled identically
to $G_{\mu}$ so that henceforth, without any loss of generality, one
of these (namely $K_{\mu}$) will be dropped.

It is now possible to provide another bosonised version of the MTM by
starting from the dual theory (48). Instead of integrating out the
Kalb-Ramond field which led to (50), one integrates over the
$B_{\mu}$-field. The result is,
\begin{eqnarray}
Z &= &\int d A_{\mu_{1}....\mu_{D-2}}
\delta(\partial_{\mu_{1}}A^{\mu_{1}....\mu_{D-2}}) exp\ i \int d^{D}x
[\frac{(1)^{D}}{2(D-1)} F_{\mu_{1}\mu_{2}...\mu_{D-1}}
F^{\mu_{1}....\mu_{D-1}} \nonumber \\
  &  &+ \frac{(-1)^{D}(D-2)!}{4n_{D}\lambda^{2}}
  A_{\mu_{1}....\mu_{D-2}} A^{\mu_{1}....\mu_{D-2}} +
  \epsilon_{\mu\nu\mu_{1}...\mu_{D-2}}
  J^{\mu}\partial^{\nu}A^{\mu_{1}....\mu_{D-2}}]
\end{eqnarray}
where the delta function prevents the occurrence of any non-local
terms. As usual, the delta function may now be expressed by its
Fourier transform and new fields $A^{'}_{\mu_{1}....\mu_{D-2}}
\rightarrow A_{\mu_{1}....\mu_{D-2}} + \partial_{[\mu_{1}}
\Lambda_{\mu_{2}....\mu_{D-2}]}$ are introduced where
$\Lambda_{\mu_{2}....\mu_{D-2}}$ is the Fourier variable. The
integration over this variable, which decouples from the primed field,
yields a trivial normalisation so that the final result simplifies to,
\begin{eqnarray}
Z &= &\int d A^{'}_{\mu_{1}....\mu_{D-2}} exp\ i \int d^{D}x
[\frac{(-1)^{D}}{2(D-1)} F_{\mu_{1}....\mu_{D-1}}(A{'})
F^{\mu_{1}....\mu_{D-1}}(A^{'}) \nonumber \\
  &  &+ \frac{(-1)^{D}(D-2)!}{4n_{D}\lambda^{2}}
  A^{'}_{\mu_{1}....\mu_{D-2}} A^{'\mu_{1}....\mu_{D-2}} +
  \epsilon_{\mu\nu\mu_{1}....\mu_{D-2}}
  J^{\mu}\partial^{\nu}A^{'\mu_{1}....\mu_{D-2}}]
\end{eqnarray}

This shows that the MTM gets identified with the theory of a massive
(D-2) rank Kalb-Ramond field. Furthermore, comparing the source terms
in (13) and (60) gives the bosonisation rule for the Thirring current,
\begin{equation}
\lambda j_{\mu} \leftrightarrow \epsilon_{\mu\nu\mu_{1}....\mu_{D-2}}
\partial^{\nu} A^{'\mu_{1}....\mu_{D-2}}
\end{equation}

An important consequence of the above analysis is that, since (50) and
(60) have a common origin (namely, the MTM in the leading $m^{-1}$
expansion), the partition functions represented by them must be
exactly equivalent. This establishes the duality between the Proca
model and the Kalb-Ramond model involving a massive (D-2) rank
antisymmetric gauge field. The basic fields in these models are
related by,
\begin{equation}
(D-2)! G_{\mu} \leftrightarrow
\epsilon_{\mu\nu\mu_{1}....\mu_{D-2}} \partial^{\nu}
A^{'\mu_{1}....\mu_{D-2}}
\end{equation}
Incidentally these findings may be regarded as the generalisation
of the D = 3 case where a self dual model is identified with the
Maxwell-Chern-Simons theory [9-11]. For the particular case of D = 4
only, a mapping similar to the one discussed here was reported
elsewhere, though in a different context [17].

It is instructive to make a counting of the number of degrees of
freedom in the two theories. The number of independent degrees of
freedom for the Proca field in D dimensions is (D-1). Now the
number of components of a (D-2) rank antisymmetric tensor in D
dimensions is $\frac{D(D-1)}{2}$. Because of constraints all of
these components are not independent. It is simple to verify that
there are (D-1)(D-2) second class (phase-space) constriants. Hence
the number of independent degrees of freedom is
$\frac{1}{2} [D(D-1) - (D-1)(D-2)] = D-1$ and agrees with the number
in the Proca theory.

\noindent {\bf Schwinger terms}

It is well known that a conventional way [14] to discuss bosonisation
is to start from Schwinger terms in the algebra of fermionic
currents. Since Schwinger terms are difficult to compute in higher
dimensions it reveals one of the reasons why bosonisation in these
dimensions, by conventional techniques, becomes problematic. Here I
shall adopt the reverse route whereby, knowing the bosonisation
relations in arbitrary dimensions, the Schwinger terms in the
current algebra \footnote{This algebra is always implied at equal
time} of fermionic theories will be evluated.

The D = 2 example is simple. Using (31) the current algebra in the
Thirring model follows trivially,
\begin{eqnarray}
i[j_{0}(x), j_{0}(y)] &= &0\nonumber\\[.5cm]
i[j_{0}(x), j_{1}(y)] &= &- \frac{1}{\pi}\partial_{1}\delta(x-y)
\end{eqnarray}
as a consequence of the equal time algebra of the free scalar
field.

The D = 2 example is special since bosonisation is exact. In higher
dimensions bosonisation has been achieved for massive fermionic
theories only in the $m^{-1}$ approximation. In the subsequent
analysis, moreover, I concentrate only on the leading order
approximation. In D = 3 dimensions the MTM is mapped on to the MCS
theory (56) with the identification (57). The components of the
current correspond to the electric and magnetic fields in the MCS
theory. Since the Poisson algebra among the latter fields is
known [11], it is straightforward to compute the current algebra,
\begin{eqnarray}
i[j_{0}(x), j_{0}(y)] &= &0 \nonumber \\
i[j_{0}(x), j_{i}(y)] &= &-
\frac{1}{\lambda^{2}}\partial_{i}\delta(x-y) \nonumber \\
i[j_{l}(x), j_{m}(y)] &=
&\frac{4\pi}{\lambda^{4}}\epsilon_{lm}\delta(x-y)
\end{eqnarray}
and agrees with the result derived previously in [10].

The last part is devoted to the $D \geq 4$ case. Here the bosonised
form of the MTM corresponds to the Proca theory (50) with the
mapping (51). The current algebra is, therefore, obtainable by the
algebra of the basic fields $G_\mu$ in the Proca theory. Since the Proca
theory is a second class system, the Poisson algebra gets replaced
by the Dirac algebra [18],
\begin{eqnarray}
i[G_{0}(x), G_{0}(y)]_{DB} &= &0 \nonumber \\[.5cm]
i[G_{0}(x), G_{i}(y)]_{DB} &= &- \frac{1}{(D-2)!}\partial_{i}\delta(x-y)
\end{eqnarray}
where the suffix `DB' denotes Dirac brackets. Using (51) and (65),
the relevant current algebra follows,
\begin{eqnarray}
i[j_{0}(x), j_{0}(y)] &= &0 \nonumber \\[.5cm]
i[j_{0}(x), j_{i}(y)] &= &- \frac{(D-2)!}{\lambda^{2}}
\partial_{i}\delta(x-y)
\end{eqnarray}
Equations (63), (64) and (66) reveal that the Schwinger terms have
an identical structure.

\section{Conclusions}
A systematic way of bosonising fermionic models in arbitrary
dimensions has been developed. The basic idea was to use gauge
invariance as a guiding principle to construct a parent lagrangian
comprising both fermionic and bosonic fields. The partition
function for this parent lagrangian in the presence of external sources
was constructed with the measure suitably modified by gauge fixing
delta functions. The bosonic integration was exactly carried out
leading to some fermionic theory like the massive Thirring model (MTM)
or the massive Schwinger model (MSM). Incidentally the structure of
the fermionic models were identical to what one would expect by
performing a duality transformation of the parent lagrangian on the
classical level. Now instead of integrating out the bosonic fields, if
the fermionic fields were integrated, the bosonised versions of the
fermionic models were obtained. Furthermore, by comparing the source
terms, explicit bosonisation identities for the fermionic currents were
derived. The fermionic integration really implies the evaluation of
the fermion determinant. Apart from D = 2 dimensional case, this
result is not exactly known. It can be computed only in some
approximation as for example the $m^{-1}$ expansion, where m is the
fermion mass. Thus the bosonisation as well as the mappings between
the various operators were valid for any $D \geq 3$ only in the large m
expansion. Moreover, contrary to other approaches [7-9] which
discuss the bosonisation of only the free theory or that in the
presence  of an external field, the analysis
presented here is valid to all orders in this inverse mass expansion.

It was shown in $D \geq 3$ that the partition functions for the MTM
and MSM can be mapped on
to the partition functions for a gauge theory involving two gauge
fields - a usual vector field and an antisymmetric (D - 2) rank
tensor field which is also called the Kalb-Ramond field. The
fermion current, likewise, gets identified with the topological
currents in the gauge theory. Furthermore, integrating over the
Kalb-Ramond field resulted in an equivalent
bosonised expression involving only the vector potential.
All these computations were valid to arbitrary orders in the inverse
fermi mass.
Simplifying the computations to the leading $(0(m^{-1}))$ order
only, it was possible instead to perform the integration over the
vector potential which led to a bosonised theory involving only
the Kalb Ramond field. In this way it was found that, in the leading
order, the MTM in D = 3 dimensions bosonises to the self dual model [12]
or to the equivalent Maxwell-Chern-Simons theory [13]. Similar conclusions
were valid for the (D = 3 dimensional) MSM. For $D \geq 4$, on the
other hand, the MTM was mapped on to the Proca model or its dual, which
was found to be the theory of a massive (D-2) rank antisymmetric tensor field.
The
bosonisation technique developed here therefore provided an elegant
way of discussing the duality between different  bosonic theories.
Incidentally, the duality between the self dual model and the
Maxwell-Chern-Simons theory in D = 3 or that between the Proca model
and the second rank massive Kalb Ramond field in D = 4 are well known
results [9-11, 17]. The present approach extends this duality for any
D-dimension. Another application was to derive the Schwinger terms in
the commutator algebra of fermionic currents by using the bosonisation
relations. Moreover all the usual results in $D=2$ bosonisation were
easily reproduced.

To put this work into proper perspective, a comparison with other
recent approaches to bosonisation is relevant. The `smooth bosonisation'
approach [3, 4], which is primarily geared for D = 2 dimensions, uses
different gauge fixing conditions in an embedded theory to recover
either the original fermionic theory or its bosonised form. The
present analysis, on the contrary, uses the same gauge condition but
reverses the order of integration in the parent theory (which
mimics the role of the embedded theory) to reproduce
once, the fermionic theory or, alternatively, the corresponding
bosonic theory. This is somewhat akin to the approach by using duality
transformations [7, 19]. However, the full gauge freedom is not exploited
there [7] leading to nonlocal terms in higher dimensional bosonisation.
Such terms never occur in my analysis. Finally, the D = 3 dimensional
appraoch  of [9] uses as a parent lagrangian the expression suggested
in [11] which is suitable for discussing bosonisation only in the
leading  $(m^{-1})$ order expansion. Significantly, all
these approaches including the present one rely on the expression
of the fermion determinant evaluated in the large-m approximation.
It is clear, therefore, that
different routes to bosonisation are connected - a fact which has
also been realised in [8] - and this analysis provided fresh insights
into this connection, apart from yielding new results in higher
dimensions.

\end{document}